\documentclass[a4paper]{article}
\pdfoutput=1
\usepackage{hyperref}

\usepackage{INTERSPEECH2021}
\hypersetup{
    colorlinks=true,
    linkcolor=red,
    filecolor=red,      
    urlcolor=red,
    citecolor=green,
}
\title{MIMO Self-attentive RNN Beamformer for Multi-speaker Speech Separation}
\name{Xiyun Li$^{1,2,3,*}$\thanks{*This work was done when Xiyun Li was an intern in Tencent.}, Yong Xu$^{3}$, Meng Yu$^{3}$, Shi-Xiong Zhang$^{3}$, Jiaming Xu$^{1,2}$, Bo Xu$^{1,2}$, Dong Yu$^{3}$}
\address{
   $^1$Institute of Automation, Chinese Academy of Sciences (CASIA), Beijing, China~~~\\
  $^{2}$School of Future Technology, University of Chinese Academy of Sciences, Beijing, China~~~\\
  $^{3}$Tencent AI Lab, Bellevue, USA~~~
  }
\email{\{lucayongxu,raymondmyu,auszhang,dyu\}@tencent.com, \{lixiyun2020,jiaming.xu,xubo\}@ia.ac.cn}

\begin{document}
\maketitle
\begin{abstract}
%   Many purely neural network (NN) based speech separation and enhancement methods can achieve good objective scores. These models , however, are harmful for the automatic speech recognition (ASR) because these models inevitably introduce nonlinear speech distortions. 
%   The minimum variance distortionless response (MVDR) beamformer with NN-predicted masks can significantly reduce speech nolinear distortions. However, the MVDR method is not optimal because it has limited noise reduction capability. 
%   On the other hand, the unstable matrix inversion and eigenvalue decomposition processes prevent these systems from being jointly trained with neural networks.
Recently, our proposed recurrent neural network (RNN) based all deep learning minimum variance distortionless response (ADL-MVDR) beamformer method yielded superior performance over the conventional MVDR by replacing the matrix inversion and eigenvalue decomposition with two RNNs. 
%In this work, we propose a self-attentive RNN beamformer to further improve the performance. 
%However, only one target speaker's speech is extracted from the multi-talker overlapped mixture in our previous work. To enable parallelization of computations and improve the inference efficiency, a unified multi-channel input and multi-target speakers' speech separation output (MIMO) model is developed in this work.
%ADL-MVDR, however, can not allow parallelization of computations and focus on multi targets speakers. 
%The self-attention mechanism, with its strong modeling capacity, is emerging as a mainstream architecture in speech. 
In this work, we present a self-attentive RNN beamformer to further improve our previous RNN-based beamformer by leveraging on the powerful modeling capability of self-attention. Temporal-spatial self-attention module is proposed to better learn the beamforming weights from the speech and noise spatial covariance matrices. The temporal self-attention module could help RNN to learn global statistics of covariance matrices. The spatial self-attention module is designed to attend on the cross-channel correlation in the covariance matrices. 
Furthermore, a multi-channel input with multi-speaker directional features and multi-speaker speech separation outputs (MIMO) model is developed to improve the inference efficiency. 
%We propose several novel MIMO self-attentive RNN beamforming methods with complex ratio filters based on the proposed self-attention modules. 
%The proposed models extend the original ADL-MVDR method to deal with multi-input and multi-output (MIMO) and accelerate the inference efficiency for multiple speakers' separation. 
The evaluations demonstrate that our proposed MIMO self-attentive RNN beamformer improves both the automatic speech recognition (ASR) accuracy and the perceptual estimation of speech quality (PESQ) against prior arts.
\end{abstract}\noindent\textbf{Index Terms}: Speech separation, MIMO, MVDR, Self-attentive RNN beamformer, spatial self-attention

\section{Introduction}

Minimum variance distortionless response (MVDR) is a widely used beamformer for the automatic speech recognition (ASR) \cite{heymann2016neural}. Recently, the mask-based MVDR \cite{heymann2016neural,boeddeker2018exploring,xiao2017time, erdogan2016improved, heymann2017beamnet,xu2020neural} achieved better ASR accuracy than purely ``black box'' neural network-based approaches \cite{hershey2016deep,yu2017permutation,chen2017deep,luo2019conv,luo2020dual} due to less non-linear distortion. However, the residual noise level of the MVDR separated speech is still high \cite{habets2013two}. Furthermore, the matrix inversion involved in the traditional MVDR solution always has the numerical instability problem \cite{zhang2021end, lim2017numerical,zhao2012fast, chakrabarty2015numerical}. Although some techniques, e.g., diagonal loading \cite{zhang2021end}, could be used to alleviate this issue, it has not been fully solved. %Some recent work investigate Time-varying beamformers  \cite{wang2019sequential,kubo2019mask}, however they still had the instability problem. 

The RNN was once demonstrated to be able to implement the matrix inversion and eigenvalue decomposition \cite{zhang2005design, wang1993recurrent, liu2008recurrent, wang2016recurrent}. Based on this, we recently proposed the ADL-MVDR \cite{zhang2020adl, zhang2020multi} beamforming method, 
%For example, our new work \cite{zhang2020adl} proposed a new method named ADL-MVDR that was a novel all deep learning MVDR framework that can be jointly trained stably with the front-end filter estimator for frame-level beamforming weights estimation. In this work, 
where two RNNs are used to replace the matrix inversion and principal component analysis (PCA) operations of the original MVDR algorithm. Compared to the conventional MVDR, better residual noise reduction and ASR accuracy are obtained. 
%from target noise and speech covariance matrices which can reduce residual noise significantly and achieve better speech recognition accuracy.
%The model can further stabilize the joint training process and estimate the covariance matrices accurately by using neural network to learn the matrix inversion and PCA from target noise and speech covariance matrices.
%In the subsequent work \cite{zhang2020multi},a generalized all deep learning MVDR framework was proposed that was capable of performing speech separation under different microphone setting, this work verified the idea of applying the ADL-MVDR framework to Multi-frame speech separation tasks and adapted the ADL-MVDR framework to a Multi-Channel Multi-frame speech separation task.
Unlike the ADL-MVDR, a more generalized RNN-based beamformer (GRNN-BF) \cite{xu2021generalized} was later proposed to learn the frame-level beamforming weights from the estimated speech and noise covariance matrices directly. The GRNN-BF can converge to a better beamforming solution without following any format of traditional beamformers (e.g, MVDR). 
%After that, our work  \cite{xu2021generalized} proved that the calculated speech and noise covariance matrices were the key factors that paved the road to design more optimal deep learning based beamformers. Following this, we seek to find a better model to calculate speech and noise covariance matrices.

% On the other hand, the transformer structure \cite{vaswani2017attention} and the self-attention \cite{shaw2018self, koizumi2020speech} scheme are popular methods for the speech separation task \cite{subakan2020attention, chen2020continuous, chen2020dual,chen2020don,tan2020sagrnn}.

%For example, Dual-path Transformer Network \cite{chen2020dual} introduces direct context-aware modeling into speech separation which can model the speech sequences directly conditioning on the context. Chen et al.  \cite{chen2020don} also applied transformer to multi-channel speech separation task.

% However, in those works, the transformer is just used for predicting the separation mask while we aim to learn the beamforming weights in this paper.

%, they propose an adaptive early exit mechanism because there doesn't need such a heavy structure in multi-channel speech separation.
%On the one hand, RNN can't finish an effective parallelization of the computations because of the inherently sequential nature of RNN models. This bottleneck is obvious in the large datasets with long sequences tasks. Transformer model can alleviate this problem. On the other hand, 
On the other hand, the self-attention \cite{vaswani2017attention,shaw2018self} is a popular method to learn the global dependencies for the speech separation task \cite{koizumi2020speech,subakan2020attention, chen2020continuous, chen2020dual,chen2020don,tan2020sagrnn,tolooshams2020channel}. However, in those works, the self-attention is used for predicting the separation mask while we aim to learn the beamforming filter from the covariance matrices in this paper. Two types of self-attention modules with different purposes are proposed to better learn the cross-channel spatial correlations and improve the temporal modeling capability of the RNN-based beamformer \cite{xu2021generalized,zhang2020adl}.

%the core module in the transformer structure \cite{vaswani2017attention} for learning global dependencies, e.g., exploiting the cross-channel correlations and cross-frame information. The transformer structure \cite{vaswani2017attention} and the self-attention \cite{shaw2018self, koizumi2020speech} scheme are popular methods for the speech separation task \cite{subakan2020attention, chen2020continuous, chen2020dual,chen2020don,tan2020sagrnn,tolooshams2020channel}. However, in those works, the transformer is just used for predicting the separation mask while we aim to learn the beamforming weights in this paper.
%more easily by establishing a direct connection. We think the self-attention layer plays an important role in transformer,  so in this work, we introduce self-attention into our model.

In this work, a MIMO self-attentive RNN beamformer is proposed to learn the frame-level beamforming weights for all speakers from all the speech and noise covariance matrices directly. There are three main contributions. First, we propose a temporal-spatial self-attention module to learn the beamforming weights. The spatial self-attention can learn the cross-channel correlations from the covariance matrices. The temporal self-attention is designed to consolidate the RNNs to capture the long-term statistics of covariance matrices. Second, the temporal self-attention, the spatial self-attention, and the RNN are demonstrated to be complementary to each other. Better performance is achieved by comparing with the RNN-based beamformer baseline  \cite{xu2021generalized, zhang2020adl}. Finally, unlike our previous target speech separation  \cite{zhang2020adl,xu2021generalized,xu2020neural}, our proposed model here is a MIMO model to enable the inference computation efficiency. It means that multi-speaker speech separation outputs could be obtained simultaneously by feeding with multiple speaker-dependent directional features.
The rest of the paper is organized as follows. In Sec. \ref{sec:cm_mtmvdr}, we describe our proposed MIMO self-attentive RNN beamformer with complex-valued ratio filter (cRF). The cRF estimator and the experimental setup are given in Sec. \ref{sec:exp}. The results are presented in Sec. \ref{sec:result}. We conclude the paper in Sec. \ref{sec:conclude}.

\begin{figure*}[htb]
\centering
	\begin{minipage}[b]{0.90\linewidth}
		\centering
		\centerline{\includegraphics[width=1.0\textwidth]{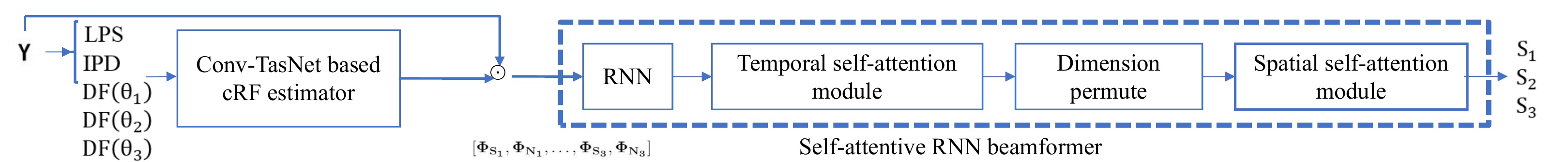}}
		%  \vspace{2.0cm}
		%\centerline{(a) Result 1}\medskip
	\end{minipage}
	\caption{The system framework consists of a complex-valued ratio filter (cRF) estimator and the proposed self-attentive RNN beamformer.
	%network structure of proposed MIMO RNN + temporal-spatial self-attentive Beamformer system. The proposed MIMO temporal-spatial self-attentive RNN Beamformer is highlighted in the dashed blue box. 
	Different from the original Conv-TasNet \cite{luo2019conv}, the encoder of our cRF estimator is a fixed STFT layer \cite{gu2020multi}. The estimated speech and noise cRFs could be used to calculate the covariance matrices.
	%The outputs of cRF estimator are all speakers' speech masks and noise masks. 
	The input of the self-attentive RNN beamformer is concatenation of all speakers' speech and noise covariance matrices. The temporal self-attention module calculates the attention matrix among frames, while the spatial self-attention calculates the attention matrix among microphone channels.} %The front-end mask estimator and the back-end acoustic model are first separately trained. Then the joint training is performed by concatenating them together.}% with the beamforming module. }
	\label{the proposed self-attentive RNN-based beamformer}
\end{figure*}

\begin{figure*}[htb]
\centering
	\begin{minipage}[b]{0.90\linewidth}
		\centering
		\centerline{\includegraphics[width=1.0\textwidth]{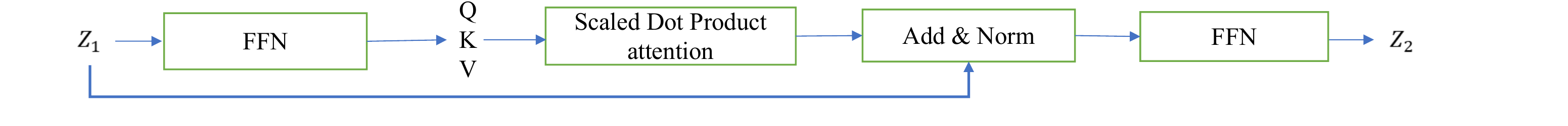}}
		%  \vspace{2.0cm}
		%\centerline{(a) Result 1}\medskip
	\end{minipage}
	\caption{The structure of self-attention module. Note that the temporal self-attention and the spatial self-attention attend on the time dimension and the spatial channel dimension, respectively.} %The front-end mask estimator and the back-end acoustic model are first separately trained. Then the joint training is performed by concatenating them together.}% with the beamforming module. }
	\label{the self-attention module}
\end{figure*}

\section{MIMO self-attentive RNN beamformer}\label{sec:cm_mtmvdr}
\subsection{Problem definition}
Given the $M$-channel mixture signal $\mathbf{y}=[\mathbf{y}_1,...,\mathbf{y}_M]$, the corresponding signal model in the short-time Fourier transform (STFT) domain is defined as,
\begin{align} \footnotesize
% \mathbf{y}(t,f)=\mathbf{x}(t,f)+\mathbf{n}(t,f),\mathbf{y}(t)=\sum_{i=1}^{C} \mathbf{s_{i}}(t)
\mathbf{Y}(t,f)&=\sum_{i=1}^{C}\mathbf{S}_{i}(t,f)+\mathbf{U}(t,f)
\label{mvdr1}
\\
% \mathbf{y}(t,f)=\mathbf{x}(t,f)+\mathbf{n}(t,f),\mathbf{y}(t)=\sum_{i=1}^{C} \mathbf{s_{i}}(t)
\mathbf{N}_{i}(t,f)&=\sum_{j\neq i}^{C}\mathbf{S}_{j}(t,f)+\mathbf{U}(t,f)
\label{i_noise}
\end{align}
where $\mathbf{S}_{i}$ and $\mathbf{U}$ represent the $i$-th speaker's reverberated speech and background noise, respectively. $\mathbf{N}_{i}$ is the corresponding interfering noise (the sum of other speakers' interfering speech and the background noise) of the $i$-th speaker. $C$ is the total number of overlapped speakers in the utterance. $t$ and $f$ are the time frame index and the frequency bin index, respectively.

Our task aims at separating all of the speakers' speech $\mathbf{S}_{i}$ simultaneously. 
% As shown in Fig. \ref{the mixture input}, the sample mixture signal includes three speakers' speech. 
Different from our previous works \cite{xu2021generalized,zhang2020adl} which focus only on separating the target speaker's speech, here we need to separate all speakers' speech simultaneously. The input includes the log-power spectra (LPS), the interaural phase difference (IPD), and multiple speakers' direction of arrival (DOA) $\theta$. Then the speaker-dependent directional feature ($\text{DF}(\theta)$) \cite{chen2018multi}, could be calculated based on the DOAs. With the speaker-dependent DOA information, the model could easily figure out the order of the separation outputs and avoid the speaker permutation problem \cite{yu2017permutation, qian2018single}. As shown in Fig. \ref{the proposed self-attentive RNN-based beamformer}, the whole system consists of two parts, a complex-valued ratio filter (cRF) \cite{mack2019deep,zhang2020adl} estimator and the proposed self-attentive RNN beamformer. The details of  the cRF estimator, which can estimate the speech and noise, will be described in Sec. \ref{sec:exp}. The proposed self-attentive RNN beamformer, which can learn the beamforming weights from the covariance matrices directly, will be introduced first here. 

\subsection{Generalized RNN-beamformer baseline} \label{sec:grnnbf_baseline}
% In our recent work \cite{zhang2020adl}, two recurrent neural networks are used to replace the matrix inversion and PCA for the estimation of the steering vector and the inverse of the noise covariance matrix for frame-level beamforming weights, which can be formulated as
% \begin{equation}
% \begin{aligned}
% \hat{\boldsymbol{v}}(t, f) &=\mathbf{G} \mathbf{R} \mathbf{U}-\mathbf{N e t}_{\boldsymbol{v}}\left(\boldsymbol{\Phi}_{\mathbf{S}}(t, f)\right) \\
% \hat{\boldsymbol{\Phi}}_{\mathbf{N}}^{-1}(t, f) &=\mathbf{G R U - N e t}_{\mathbf{N}}\left(\mathbf{\Phi}_{\mathbf{N}}(t, f)\right) \\
% \mathbf{h}(t, f)&=\frac{\hat{\mathbf{\Phi}}_{\mathbf{N}}^{-1}(t, f) \hat{\boldsymbol{v}}(t, f)}{\hat{\boldsymbol{v}}^{\mathrm{H}}(t, f) \hat{\mathbf{\Phi}}_{\mathbf{N}}^{-1}(t, f) \hat{\boldsymbol{v}}(t, f)}, \quad \mathbf{h}(t, f) \in \mathbb{C}^{M}
% \end{aligned}
% \end{equation}

% $v(f)$ denotes the steering vector of the target speech,
% the real and imaginary parts of the complex-valued covariance matrix are concatenated together as input to the GRU-Net. The model can accumulate and update the covariance matrix for each frame leveraging the temporal structure of RNNs.

Our recent work \cite{xu2021generalized} proposed a generalized RNN-based beamformer (GRNN-BF) method. The GRNN-BF used the complex-valued ratio filter (cRF) \cite{mack2019deep} to calculate the target speech and noise covariance matrices. The cRF is just an extended version of complex-valued ratio mask (cRM) \cite{williamson2015complex} by using the neighboring context information. Then a RNN model was applied to learn the frame-level beamforming weights directly from the covariance matrices. 
%$(2K+1)\times(2K+1)$ neighboring context was used to stabilize the calculation of covariance matrices.
As shown in Fig. \ref{the proposed self-attentive RNN-based beamformer}, the cRF estimator first predicts the target speech and noise cRFs. Then the estimated $i$-th speaker's speech $\hat{\mathbf{S}}_{i}(t, f)$ is,
\begin{equation} \footnotesize
\hat{\mathbf{S}}_{i}(t, f)
=\sum_{\tau_{1}=-L}^{\tau_{1}=L} \sum_{\tau_{2}=-L}^{\tau_{2}=L} \operatorname{cRF}_{\mathbf{S}_{i}}\left(t+\tau_{1},f+\tau_{2}\right)*\mathbf{Y}\left(t+\tau_{1},f+\tau_{2}\right)\end{equation}
where $L$ defines the neighboring context size across the frequency bins and the time frames. The corresponding $i$-th speaker's noise $\hat{\mathbf{N}}_{i}(t,f)$ with the corresponding noise $\text{cRF}_{\mathbf{N}_{i}}(t,f)$ could be estimated in the same way. The frame-wise $i$-th speaker's speech covariance is calculated as,
% \begin{equation}  \footnotesize 
% \label{eq:cov_matrix}
%     \boldsymbol{\Phi}_{{\mathbf{S}}_{i}}(t, f)=\frac{\hat{\mathbf{S}}_{i}(t, f) \hat{\mathbf{S}}_{i}^{\mathrm{H}}(t, f)}{\sum_{t=1}^{T} \operatorname{cRM_{S_i}}^{\mathrm{H}}(t, f) \operatorname{cRM_{S_i}}(t, f)}
% \end{equation}
\begin{equation}  \footnotesize 
\label{eq:cov_matrix}
    \boldsymbol{\Phi}_{{\mathbf{S}}_{i}}(t, f)=\frac{\hat{\mathbf{S}}_{i}(t, f) \hat{\mathbf{S}}_{i}^{\mathrm{H}}(t, f)}{\sum_{t=1}^{T} \operatorname{cRM}_{\mathbf{S}_{i}}^{\mathrm{H}}(t, f) \operatorname{cRM}_{\mathbf{S}_{i}}(t, f)}
\end{equation}
where $\operatorname{cRM}_{\mathbf{S}_i}(t,f)$ stands for the center mask of the $\operatorname{cRF}_{\mathbf{S}_i}(t,f)$. Then RNNs are used to directly learn the frame-level beamforming weights from the frame-wise covariance matrices \cite{xu2021generalized}, which can be formulated as 
\begin{align}  \footnotesize
\mathbf{I}_\text{i}(t,f) &= [{{\bf{ \Phi}}_{\mathbf{S}_{i}}(t,f), \bf{{ \Phi}}}_{\mathbf{N}_{i}}(t,f)] \label{eq:I_cat}\\
\mathbf{w}_\text{1}(t,f),..., \mathbf{w}_\text{C}(t,f) &= \text{RNN}([{\mathbf{ I}}_{1}(t,f), ..., {\mathbf{I}_{C}}(t,f)])
\end{align}
where $\mathbf{I}_\text{i}(t,f)$ stands for the concatenation of $i$-th speaker's speech and noise covariance matrices, $\mathbf{w}_\text{i}(t,f) \in \mathbb{C}^{M}$ denotes the $\text{i}$-th speaker's beamforming weights. 
\subsection{Proposed MIMO self-attentive RNN beamformer}
Fig. \ref{the proposed self-attentive RNN-based beamformer} shows the detailed architecture of our MIMO self-attentive RNN beamforming system. 
%Fig .\ref{the self-attention module} shows the self-attention module. 
The model consists of a cRF estimator and the proposed self-attentive RNN beamformer.
% RNN-temporal-spatial self-attention Beamforming method. 
%Fig .\ref{the self-attention module} shows the detailed information of our self-attention module. 
% Similar to the ADL-MVDR method, our model consists of a complex filter mask estimator which is based on our previously proposed multi-modal MC speech separation platform \cite{tan2019audio,lorry2020}(a variant of Conv-TasNet \cite{luo2019conv} to estimate frame-level covariance matrices) and proposed self-attentive RNN beamformer for frame-level beamforming weights derivation.
% that employs several recurrent neural networks and self-attention modules to predict frame-wise beamforming weights
% . On the one hand, using RNN can utilize the temporal information from all previous frames. On the other hand, self-attention can utilize the weighted information from all frames directly. Our model estimates the speech and noise components using a complex ratio filtering method \cite{zhang2020adl} to better utilize the nearby T-F information and stabilize the estimated statistical variables.
Similar to GRNN-BF \cite{xu2021generalized}, the cRFs first help to estimate the speech and noise covariance matrices (as shown in Eq. (\ref{eq:cov_matrix})). Then the self-attentive RNN beamformer could predict the frame-level beamforming weights from the covariance matrices. 
% the self-attentive RNN-beamforming module 
% as shown in Fig .\ref{the proposed self-attentive RNN-based beamformer} 
%the input of our proposed self-attentive module is a three-dimensional tensor comprising of concatenation of all the target speakers' noise and speech covariance matrices. 
% We propose several self-attentive RNN beamforming methods for this task
Several types of self-attention modules, namely temporal self-attentive module, and spatial self-attentive module, temporal-spatial self-attentive module are proposed to further improve the RNN-based beamformer \cite{xu2021generalized} to learn a better beamforming filter in this work. %Fig .\ref{the self-attention module} shows the detailed steps of the self-attention module. 
\subsubsection{Temporal self-attentive module}
%Proposed MIMO temporal self-attentive RNN beamformer has a similar framework to Fig .\ref{the proposed self-attentive RNN-based beamformer} and Fig .\ref{the self-attention module}. 
Self-attention \cite{vaswani2017attention}, with the powerful modeling capability, is widely used in the speech enhancement and speech separation tasks \cite{koizumi2020speech,subakan2020attention,wang2021continuous,chen2020continuous,tan2020sagrnn}. We first propose a temporal self-attention module to improve the temporal modeling capability of RNNs. %Although RNN is a very popular model in speech-related tasks, RNN can only learn local patterns[引文？？？], and self-attention can better learn long-term dependencies and cross-frame information. Therefore, we introduce temporal self-attention module to make up for the shortcomings of RNN.
%Similar to standard self-attention \cite{vaswani2017attention}, the hidden output 
As shown in Fig. \ref{the self-attention module}, the input $\mathbf{Z}_{1}$ to the self-attention module is processed by three linear transform layers on the feature dim followed by the PReLU activation, which is denoted as the feed-forward network (FFN). The outputs of the FFN are represented as $\mathbf{Q}$, $\mathbf{K}$, $\mathbf{V}$. Then a self-attention function is applied to exploit cross-frame correlation, followed by a residual path and a layer normalization. The temporal self-attention calculates the attention matrix among frames,
% We use the variable $\mathbf{O}$ to denote ntion module is implemented as follows:
\begin{equation}  \footnotesize
\label{eq:self_att}
\begin{aligned} 
\text{Attention(}\mathbf{Q},\mathbf{K},\mathbf{V}\text{)}=\operatorname{softmax}\left(\frac{\mathbf{Q}\mathbf{K}^{\top}}{\sqrt{d_{k}}}\right) \mathbf{V} 
\end{aligned}
\end{equation}
where $d_{k}$ stands for the hidden layer dimension of $\mathbf{K}$. Finally, the output of the self-attention function is fed into another FFN to get the transformed output $\mathbf{Z}_{2}$. Different from other works \cite{koizumi2020speech,subakan2020attention,wang2021continuous,chen2020continuous,chen2020don} which apply self-attention layers to predict the separation mask, we use a self-attention module to estimate the beamforming weights.
% The complex filter mask estimator adopts a variant of Conv-TasNet architecture to estimate frame-level covariance matrices which is based on our previously proposed multi-modal MC speech separation platform \cite{tan2019audio,lorry2020}.
% The input of our proposed self-attentive module is a three-dimensional tensor comprising of concatenation of all the target speakers' noise and speech covariance matrices, followed by a DNN-RNN layer. 
%A cross-frame self-attention layer in temporal self-attention module can learn to focus on no-silent segments and consolidate the RNNs to capture the long-term statistics of covariance matrices \cite{wang2021continuous,liu2020self,zheng2020interactive}, so the self-attention layer can learn long-term pattern, which can make up for the shortcoming of RNN. 
% Similar to standard self-attention, the temporal self-attention module processes the cross-frame correlation which can learn to focus on no-silent segments and consolidate the RNNs to capture the long-term statistics of covariance matrices.
% \begin{figure}[htb]
% 	\begin{minipage}[b]{0.9\linewidth}
% 		\centering
% 		\centerline{\includegraphics[width=1.0\textwidth]{pic570.png}}
% 		%  \vspace{2.0cm}
% 		%\centerline{(a) Result 1}\medskip
% 	\end{minipage}
% 	\caption{the self-attention module} %The front-end mask estimator and the back-end acoustic model are first separately trained. Then the joint training is performed by concatenating them together.}% with the beamforming module. }
% 	\label{the self-attention module0}
% \end{figure}
\subsubsection{Spatial self-attentive module}
To better learn cross-channel correlations from the covariance matrices, we also propose a spatial self-attention module. Instead of acting on the time dimension as the temporal self-attention did, the spatial self-attention module attends on the spatial channel dimension to learn the cross-channel information. As shown in Fig. \ref{the proposed self-attentive RNN-based beamformer}, a dimension permute operation is applied before the spatial self-attention module is conducted. 
%by an additional dimension permute operation. After a RNN layer, we permute the last two dimensions and the spatial self-attention module is applied to model the cross-channel correlations.

Different from \cite{wang2021continuous} where a similar spatial self-attention module is applied on the amplitude spectrogram to learn the separation mask, our proposed self-attention module learns on the complex-valued frequency domain for predicting the beamforming filter. In the amplitude of the frequency domain, the phase information of different microphone channels is lost and the self-attention can not effectively utilize the spatial correlations. However, our proposed spatial self-attention module could fully learn the spatial correlation from the complex-valued speech and noise covariance matrices.
%use complex in frequency domain to learn cross-channel correlation instead of amplitude spectrogram.  Amplitude spectrogram \cite{wang2021continuous} is not optimal due to the loss of phase difference information among channels which is harmful for exploiting nolinear spatial correlation between channels. Furthermore, \cite{wang2021continuous} used Transformer-based spatial-temporal modeling for estimation of frequency domain mask while we apply temporal-spatial self-attention module for frame-level beamforming weights.
% RNN-spatial self-attention module that focuses on the feature dimension can learn 
% the cross-channel correlation in the covariance matrix while RNN-temporal self-attention module which focuses on the time dimension can learn to focus on no-silent segments and consolidate the RNNs to capture the long-term statistics of covariance matrices. The RNN-temporal self-attention method applies the temporal self-attention module while the RNN-spatial self-attention method uses the spatial self-attention module. RNN-temporal-spatial self-attention beamforming method combines both modules and temporal-spatial self-attention beamforming integrates both instead of the RNNs.
\subsubsection{MIMO temporal-spatial self-attentive RNN beamformer}
With the temporal self-attention module and the spatial self-attention module, we can combine them into the proposed temporal-spatial self-attentive RNN beamformer to model both cross-frame information and cross-channel correlations (as shown in Fig. \ref{the proposed self-attentive RNN-based beamformer}). 
%by combining spatial self-attention module and temporal self-attention module. 
% The input of our proposed self-attentive module is a three-dimensional tensor comprising of concatenation of the target noise and speech covariance matrices, followed by a DNN-RNN layer. 
%A cross-frame self-attention layer in temporal self-attention module can learn the long-term statistics and capture a long-range acoustic context \cite{wang2021continuous,liu2020self,zheng2020interactive}. After a dimension permute module, a spatial self-attention module exploits nolinear spatial correlation between different channels and was shown effective in  \cite{wang2020neural,wang2021continuous}.
Similar to the GRNN-BF baseline (see Sec. (\ref{sec:grnnbf_baseline})), the input to the self-attentive RNN beamformer (SA-RNN) is also the concatenated speech and noise covariance matrices of all speakers. Then the multiple speakers' beamforming weights $\mathbf{w}_{i}(t, f)  \in \mathbb{C}^{M}$ and the separated $i$-th speaker's speech $\mathbf{S}_{i}$ could be predicted as,
\begin{align}  \footnotesize
\mathbf{w}_{1}(t, f),...,\mathbf{w}_{C}(t, f)&=\text{SA-RNN}\left(\left[ {\mathbf{ I}}_{1}(t,f),..., {\mathbf{{I}}}_{C}(t,f)\right]\right) \\
%Then the separated $i$-th speaker's speech $\mathbf{S_{i}}$ is obtained as 
\mathbf{S_{i}}(t, f)&=\mathbf{w}_{i}(t, f)^{\mathrm{H}} \mathbf{Y}(t, f), i=1,\ldots,C \end{align}
%where the SA-RNN denotes Self-attentive RNN beamforming module. 
%The output of the model includes every target speaker's reverberated speech $\mathbf{S_{i}}(t),i=1,\ldots,C$. As we can see in Fig . \ref{the proposed self-attentive RNN-based beamformer} and Fig . \ref{the self-attention module}, this MIMO framework can have an effective parallelization of the computations. 
%where ${\mathbf{I}}_{i}(t,f)$ is defined in Eq. (\ref{eq:I_cat}). 
Note that this work focuses on speech separation and de-noising without de-reverberation. The scale-invariant signal-to-noise ratio (Si-SNR) \cite{luo2019conv} loss between the estimated speech and the reverberated clean speech is calculated and used to train the model in an end-to-end mode. Each speaker's Si-SNR loss is averaged if that speaker exists in the utterance.
%where $\mathbf{w_{i}(f)}$ represents the beamforming weights at frequency index f for the speaker $i$ and $^H$ stands for the Hermitian operator. 
% The self-attentive RNN model can learn long term statistics of covariances and distant relationships by establishing connections between distant elements directly, which can make up for the shortcoming of RNN. 
\section{Experimental Setup and Dataset}\label{sec:exp}
%We evaluate our proposed methods on our multi-channel target speech separation platform  \cite{tan2019audio,lorry2020}.
\subsection{System overview}
As shown in Fig. \ref{the proposed self-attentive RNN-based beamformer}, the system consists of a cRF estimator and the proposed self-attentive RNN beamformer. The cRF \cite{mack2019deep} estimator is developed based on the Conv-TasNet \cite{luo2019conv} with the fixed STFT encoder \cite{gu2020multi,tan2019audio}. %architecture which has significantly surpassed many deep learning methods in the time domain or frequency domain to estimate frame-level covariance matrices. Our complex filter mask estimator is based on our previously proposed multi-modal MC speech separation platform \cite{tan2019audio,lorry2020,bahmaninezhad2019comprehensive}.
% Our mask estimator network is based on the multi-channel  \cite{tan2019audio, bahmaninezhad2019comprehensive,lorry2020} speech separation Conv-TasNet \cite{luo2019conv} model which has significantly surpassed many deep learning methods in the time domain or frequency domain and improved performance of both objective distortion measures and subjective quality assessment. 
The inputs to the cRF estimator include the 15-channel microphone mixture and all speakers' DOA information. Specifically, the LPS of the 1st channel $\mathbf{Y}^{(0)}$, IPD pairs \cite{tan2019audio} and multiple speakers' directional features $\text{DF}(\theta)$ are concatenated into the cRF estimator. The location guided $\text{DF}(\theta)$ \cite{chen2018multi} calculates the cosine similarity between the $i$-th speaker's steering vector $v({\theta}_i)$ and IPDs \cite{chen2018multi}. With $\text{DF}(\theta)$ features, our model can extract the target speaker's speech from the specific DOA and it can avoid the speaker permutation problem \cite{yu2017permutation}. As for the simulated data, the ground-truth DOA is known. For the real-world scenario, our actual hardware including a 15-microphone non-uniform linear array aligned with a $180^{\circ}$ wide-angle camera (see our demo at \href{https://lixiyun98.github.io/SA-RNN/}{https://lixiyun98.github.io/SA-RNN/}). Then the DOA could be roughly estimated through the location of the speaker's face \cite{tan2019audio}. The predicted speech and noise cRFs are used to calculate the covariance matrices. Finally, the proposed MIMO self-attentive RNN beamformer could learn the  beamforming weights for multiple speakers from covariance matrices.
%of our model are extracted from the 15-channel microphone recorded mixture that is synchronized with the $180^{\circ}$ wide-angle camera \cite{xu2020neural}. The 15-element non-uniform linear microphone array  \cite{tan2019audio} is co-located with the $180^{\circ}$ wide-angle camera. The inputs of our model include the speaker-independent features log-power spectra (LPS) and interaural phase difference (IPD)  \cite{tan2019audio},
%speaker-dependent feature directional feature (DF  \cite{chen2018multi,lorry2020}($\theta$)). $\theta$ is the speaker-dependent direction of arrival(DOA). The LPS, IPDs and DF($\theta$) are merged and fed into a bunch of dilated 1D-CNNs. 
% AF feature can slove the permutation problem. 
%The details can be found in our previous work  \cite{tan2019audio,lorry2020}. Then the concatenated directional information and audio embeddings  \cite{tan2019audio} are used to predict the complex-ratio filter. The location guided directional feature (DF($\theta$)) \cite{chen2018multi} $d(\theta)$ aims at calculating the cosine similarity between the target steering vector $v({\theta})$ and IPDs  \cite{chen2018multi}. With DF feature, Model can extract the target speech from the specific DOA. The location of the target speaker's face in the whole camera view can provide a rough DOA estimation of the target speaker which can inform the dilated convolutional neural networks to extract the target speech from the multi-talker mixture. In this paper, the input of our model don't include the lip movement feature because we focus on the beamforming.
\subsection{Dataset and experimental setup}
\begin{table*}[htbp]
	\centering
	\caption{PESQ, Si-SNR and WER results of several MIMO baselines and the proposed MIMO self-attentive RNN Beamformer system.}
	% Table generated by Excel2LaTeX from sheet 'Sheet1'
	% Table generated by Excel2LaTeX from sheet 'Sheet1'
	% Table generated by Excel2LaTeX from sheet 'Sheet1'
	\begin{tabular}{l|ccc|c|c|c}
		\hline
		Systems/Metrics & \multicolumn{3}{c|}{Si-SNR (dB)} &    Si-SNR   & WER (\%) & PESQ\\
% 		\hline
% 		 & \multicolumn{3}{c|}{ \# of overlapped speakers} &         &  \\
		\hline
		& SPK1 & SPK2 & SPK3 & Ave & Ave & Ave \\
		\hline
		Reverberant Clean (reference) & $\infty$ & $\infty$ & $\infty$ & $\infty$ & 7.28\% & 4.50\\
		Mixture (interfering speech + noise) & 21.85  & 0.91  &  -0.52 & 2.47 & 80.46\% & 1.76\\
		\hline
		MIMO Conv-TasNet with STFT  \cite{gu2020multi} (i)  & 25.24  & 11.14  & 7.93  & 11.25   & 24.21\% &3.00\\
		MIMO MVDR \cite{xu2020neural} (ii) & 20.19  & 9.85  & 6.38  & 9.47 & 17.17\% & 3.03\\
		MIMO Generalized RNN (GRNN) beamformer \cite{xu2021generalized} (iii)  & 27.32 & 14.36  & 11.12  & 14.34 & 12.18\% & 3.46 \\
		\hline
		Prop. MIMO RNN + temporal SA beamformer (iv)   & 27.96  & 15.51  & 12.15  & 15.39  & 11.26\% & 3.7 \\
		Prop. MIMO RNN + spatial SA beamformer (v)  & 28.21  & 15.78  & 12.6  & 15.74   & 10.66\% & 3.73 \\
		Prop. MIMO temporal-spatial SA beamformer (vi)  & 27.61  & 15.34 & 12.03 & 15.22  & 11.08\% & 3.66 \\
		\textbf{Prop. MIMO RNN + temporal-spatial SA beamformer} (vii)  & \textbf{28.53}  & \textbf{16.65}  & \textbf{13.25}  & \textbf{16.46}  & \textbf{10.13\%} & \textbf{3.78} \\
		\hline
	\end{tabular}%
	\label{tab:pesq}%
\end{table*}%
%The new and larger multi-talker multi-channel far-field is collected from Youtube. We use SNR estimation tool to filter out low SNR ($\le$ 17dB)  \cite{tan2019audio},
The methods are evaluated on the mandarin audio-visual corpus \cite{tan2019audio,gu2020multi}, which is collected from YouTube \cite{zhang2020multi} with about 200 hours over 1500 speakers. 
%The multi-channel multi-talker dataset is simulated in a similar way with our previous work \cite{tan2019audio,lorry2020}.
The multi-channel multi-talker simulated dataset contains 153800, 500, and 1053 multi-channel mixtures for training, validation, and testing. There are up to three overlapped speakers in one utterance. 
% Hence the scenarios are split into 1-speaker, 2-speaker, and 3-speaker. 
%The speakers in the test set are unseen in the training set. 
The multi-channel signals are generated by convolving speech with room impulse responses (RIRs) simulated by image-source method \cite{habets2006room}. The virtual acoustic room size is ranging from 4m-4m-2.5m to 10m-8m-6m. The reverberation time T60 is sampled in a range of 0.05s to 0.7s. The signal-to-interference ratio (SIR) is ranging from -6 to 6 dB. Also, noise with 18-30 dB SNR is added to all the multi-channel mixtures \cite{tan2019audio}.

In our experiment, the sampling rate for audio is 16 kHz. 512-point of STFT is used to extract audio features along with 32ms Hann window with 50\% overlap. Similar to the ADL-MVDR \cite{zhang2020adl} and GRNN-BF \cite{xu2021generalized}, the size of cRF \cite{zhang2020adl,mack2019deep} is empirically set to $3 \times 3$. The network is trained in a chunk-wise mode with chunk size of 4 seconds and batch size of 8, using Adam optimizer with early stopping. Pytorch 1.1.0 is used. The initial learning rate is set to 1e-4. Gradient norm is clipped with max norm 10. 
%The zero paddings of existed speakers need to be added with small eps 1e-6. We apply loss function Si-SNR on existed speakers including the padding zeros of existed speakers. We skip the chunks with $<1s$ valid speech for all speakers such as some chunks with $<1s$ speaker 1 or with $<1s$ speaker 2 because the model is trained in a chunk mode. Pytorch 1.1.0 was used. Our objective is to maximize the Si-SNR loss and all models are trained for 60 epochs.
In terms of the self-attentive RNN beamformer module, 
%the input consists of a concatenation of the target noise and speech covariance matrices. In our experiment, the input includes speaker 1's noise and speech covariance matrice, speaker 2's noise and speech covariance matrice, and speaker 3's noise and speech covariance matrice. 
the network consists of a 2800-size fully connected layer (FC) followed by a uni-directional gated recurrent unit (GRU) layer with PRelu activation function. The hidden size is set to 500 for the GRU layer. 
%The dimension permute is used for the Spatial self-attention module. The difference between spatial self-attention module and temporal self-attention module is the scaled dot product attention on time dim or feature dim. 
The hidden size for the output linear FC layer is 90.

A commercial general-purpose mandarin speech recognition API \cite{tencent_api} is used to test the ASR performance. The PESQ, Si-SNR \cite{luo2019conv} and ASR word error rate (WER) results are shown in Table \ref{tab:pesq}. 
%to compare among purely network-based systems, generalized RNN Beamforming \cite{xu2021generalized}, several proposed MIMO self-attentive RNN beamformers(MIMO SA beamformers) and traditional MVDR system \cite{xu2020neural}. Note that we only conduct speech separation and denoising without dereverberation in this work. Our systems work well on different scenarios, e.g., various number of overlapped speakers. The scenarios, e.g., more overlapped speakers, are a bit more challenging. 
Some real-world recording demos can be found at \href{https://lixiyun98.github.io/SA-RNN/}{https://lixiyun98.github.io/SA-RNN/}.

\section{Results and Discussions}\label{sec:result}
\begin{figure}[htb]
\centering
	\begin{minipage}[b]{\linewidth}
		\centering
		\centerline{\includegraphics[width=1.0\textwidth]{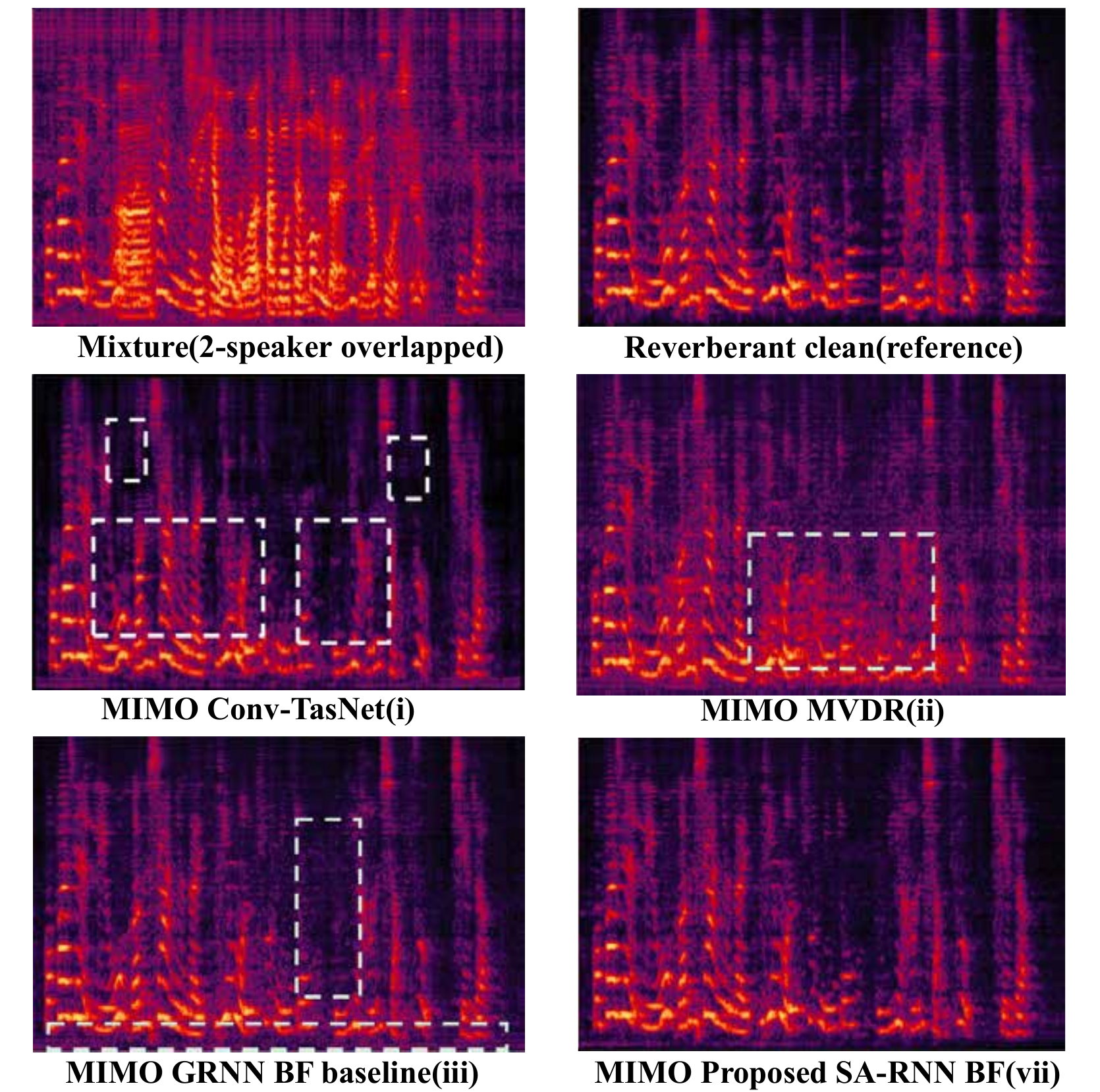}}
		%  \vspace{2.0cm}
		%\centerline{(a) Result 1}\medskip
	\end{minipage}
	\caption{Sample spectrograms of different systems. The non-linear distortion and residual noise are highlighted in the dashed white box.} %The front-end mask estimator and the back-end acoustic model are first separately trained. Then the joint training is performed by concatenating them together.}% with the beamforming module. }
	\label{Sample spectrograms of some evaluated systems}
\end{figure}
%In the previous work \cite{zhang2020adl,xu2021generalized}, under three situations (only one speaker, two speakers, and three speakers), only the performance of the 1st speaker was calculated. In our work, the 1st result represents the average Si-SNR of all the 1st speakers in all three situations, instead of the performance when only one speaker.

In Table \ref{tab:pesq}, we present the averaged speech separation performance of multiple target speakers. There might be one to three overlapped speakers in one utterance. The detailed Si-SNRs of $i$-th speaker are also presented. The order of the speaker is defined by the order of the speakers' DOAs. The SPK3 case means that there must be three overlapped speakers in one utterance and the result of SPK3 represents the performance of the third speaker in three overlapped speakers, which is the most difficult scenario for overlapped speech separation. 

%Several baselines are used for performance comparisons. The first baseline is the MIMO Conv-TasNet \cite{luo2019conv} (i) using fixed STFT encoder \cite{gu2020multi}. It is the cRF estimator without the beamformer module as shown in Fig. \ref{the proposed self-attentive RNN-based beamformer}. 

\textbf{SA-RNN beamformer vs. MVDR}: The MIMO MVDR \cite{xu2020neural} baseline (ii) also uses the cRF estimator to calculate the covariance matrices but replaces the beamforming module in Fig. \ref{the proposed self-attentive RNN-based beamformer} with the conventional MVDR solution \cite{xu2020neural}. The proposed temporal-spatial self-attentive RNN (SA-RNN) beamformer (vii) outperforms the MVDR \cite{xu2020neural} (ii) by a large margin in terms of WER, PESQ, and Si-SNR score. Significant improvements across objective metrics are observed (i.e., WER: 10.13\% vs. 17.17\%, PESQ: 3.78 vs. 3.03, Si-SNR: 16.46 dB vs. 9.47dB). 
%The proposed MIMO self-attentive RNN beamforming system (vii) achieves about 40\% improvement (i.e., Si-SNR: 16.09 dB vs. 9.47 dB) over the baseline MVDR system (ii). 
% There remains a lot of residual noise although MIMO MVDR systems(ii) can alleviate the distortion issue. 
Especially, under the extreme condition where three speakers are overlapped, our proposed system (vii) improves the Si-SNR of SPK3 from 6.38 dB to 13.25 dB. 
% The experimental results presented here verify our claims that our proposed system not only ensures the distortionless of the target speech(i.e., lowest WER) but also reduces the residual noise(i.e., highest scores across PESQ and Si-SNR).
This can also be observed in Fig. \ref{Sample spectrograms of some evaluated systems}, where lots of residual noise could be seen from the separated spectrogram of MVDR. This is because the MVDR has limited noise reduction capability \cite{habets2013two}.
% However, our MIMO MVDR (ii) works reasonably well.
%With the distortionless constraint \cite{habets2013two}, the MVDR achieves lower WER than the MIMO Conv-TasNet baseline (i).
% and be highlighted in Fig. \ref{Sample spectrograms of some evaluated systems}. 

% With the distortionless constraint in the MVDR, the beamformed speech can achieve much lower WER than our CRM baseline. For example, the jointly trained CM-based MVDR can reduce the WER from 24.21\% to 17.17\% when compared to the CM-based network w/o MVDR even CM is superior to other real-valued masks (ReLU mask or sigmoid mask) in estimating the target speech and noise covariance matrix. However, MVDR beamformer obtains this distortionless advantage by sacrificing the strength of residual noise reduction  \cite{habets2013two}, e.g., MVDR (vii) only achieves 2.91 PESQ on average and is lower than purely network-based system (iv) with 3.00 PESQ. Of course, this can be observed in the objective scores and can be highlighted in .

\textbf{SA-RNN beamformer vs. GRNN beamformer}: The MIMO generalized RNN beamformer (GRNN-BF) (iii) is described in Sec. \ref{sec:grnnbf_baseline}. 
The proposed SA-RNN beamformer (vii) achieves better performance in all metrics (e.g., Si-SNR: 16.46 dB vs. 14.34 dB, PESQ: 3.78 vs. 3.46, WER: 10.13\% vs. 12.18\%) than the GRNN-BF baseline \cite{xu2021generalized} (iii). A relative 16.8\% WER reduction is achieved. It indicates that the temporal-spatial self-attention could improve the RNN to learn a better beamforming solution from the covariance matrices by fully using the cross-frame and cross-channel correlations. Compared to the separated sample spectrogram of the GRNN-BF in Fig. \ref{Sample spectrograms of some evaluated systems}, the proposed SA-RNN beamformer (vii) could reduce more residual noise. 
%by comparing with the GRNN-BF baseline. Slight improvements can be found on the RNN beamformer since the introduction of self-attention. It indicates that the benefits of introducing a self-attention layer include further reducing the residual noise while not distorting the speech.

\textbf{SA-RNN beamformer vs. Conv-TasNet}: The MIMO Conv-TasNet (i) with a fixed STFT encoder \cite{gu2020multi} is a variant of the original Conv-TasNet \cite{luo2019conv}. It is the cRF estimator without the beamforming module as shown in Fig. \ref{the proposed self-attentive RNN-based beamformer}. The input to the Conv-TasNet is the same multi-channel information with other systems. The proposed SA-RNN beamformer (vii) achieves higher PESQ (3.78 vs 3.00), higher Si-SNR (16.46 dB vs. 11.25 dB) and lower WER (10.13\% vs 24.21\%) compared to MIMO Conv-TasNet with STFT \cite{gu2020multi} (i) baseline. For the MIMO Conv-TasNet with STFT (i), we can find that it performs the worst in the WER metric (i.e., 24.21\%) among all systems due to a large amount of non-linear distortion. The non-linear distortion in the separated speech is not ASR friendly. This non-linear distortion problem always exists in purely ``black box'' neural network based methods \cite{du2014robust,xu2020neural}. The difference between the Conv-TasNet (i) and the SA-RNN beamformer (vii) is also shown on the sample spectrograms in Fig. \ref{Sample spectrograms of some evaluated systems}.
% In terms of Si-SNR, large gaps can be found(i.e.: 15.9 dB vs. 11.25 dB).
% Significant improvements aross objective metrics are also observed. 
% This phenomenon is widely observed in purely neural network based speech enhancement front-ends  \cite{du2014robust} because the non-linear distortion in the enhanced speech is not ASR friendly. 
% Even for the commercial general-purpose ASR engine, non-distortion is more important than no residual noise, considering that the ASR engine is already robust to the mild level noise (but may not robust to the non-linear distortion). 

\textbf{Several types of SA-RNN beamformers}:
Different types of proposed SA-RNN beamformers are investigated for the ablation study. Both of the RNN with temporal only self-attention (iv) and spatial only self-attention (v) work better than the GRNN-BF baseline (iii) without self-attention. The RNN with spatial only self-attention (v) is better than the RNN with temporal only self-attention (iv), e.g., WER: 10.66\% vs. 11.26\%. However, the temporal-spatial self-attention (vi) without a RNN works worse than the best SA-RNN system (vii), e.g., WER: 10.13\% vs. 11.08\%. These comparison results suggest that the RNN, temporal self-attention, and spatial self-attention are complementary to each other. Best performance could be achieved by combining them to jointly learn the beamforming weights.
%MIMO self-attentive SA beamformers (iv,v,vi,vii) are proposed in this paper. These beamformers still follow a similar framework as we can see in Fig. \ref{the proposed self-attentive RNN-based beamformer} and Fig. \ref{the self-attention module}. 
%The RNN-SA Beamformer (vii) achieves the best performance in terms of PESQ, Si-SNR and WER.  Compared to the GRNN-BF (iii), the RNN + temporal-spatial self-attention Beamformer (vii) can obtain better performance with 0.28 higher PESQ and relatively 2\% WER reduction. It indicates that by combining the temporal self-attention module, the spatial self-attention module, and RNN we can significantly obtain the optimal beamforming weights for each frame and better noise reduction capability.

\section{Conclusions}\label{sec:conclude}
In this work, we proposed a MIMO self-attentive RNN beamformer, including a temporal self-attention module and a spatial self-attention module. Better beamforming weights are learned and higher inference efficiency could be achieved with the introduced self-attention and MIMO schemes. Compared to the prior art method, GRNN-BF \cite{xu2021generalized}, a relative 16.8\% WER reduction and a relative 14.8\% Si-SNR improvement could be achieved. In future work, we will further introduce multi-head attention based Transformer \cite{vaswani2017attention} into our model.
%and move on to the dereverberation task.
% $$\mathbf{cRF_{S_{1}}},\mathbf{cRF_{S_{2}}},\mathbf{cRF_{S_{3}}},\mathbf{cRF_{N_{1}}},\mathbf{cRF_{N_{2}}},\mathbf{cRF_{N_{3}}}$$
% $$\mathbf{cRFs}$$
% $$\mathbf{\Phi}$$
%The experimental results presented here verify our claims that the proposed self-attentive RNN beamformer not only ensures the distortionless of the target speech (i.e., lowest WER ) but also eliminates the residual noise (i.e., highest scores across all objective metrics). Compared to the Generalized RNN beamforming \cite{zhang2020adl,zhang2020multi,xu2021generalized}, our model can significantly reduce the WER from 24.21 \% to 10.34\% and improve the PESQ from 3.00 to 3.76 on average. We will further introduce multi-heads multi-layers transformer into our model and move on to the dereverberation task.

\bibliographystyle{IEEEtran}
\bibliography{mybib}
% \begin{thebibliography}{9}
% \bibitem[1]{Davis80-COP}
%   S.\ B.\ Davis and P.\ Mermelstein,
%   ``Comparison of parametric representation for monosyllabic word recognition in continuously spoken sentences,''
%   \textit{IEEE Transactions on Acoustics, Speech and Signal Processing}, vol.~28, no.~4, pp.~357--366, 1980.
% \bibitem[2]{Rabiner89-ATO}
%   L.\ R.\ Rabiner,
%   ``A tutorial on hidden Markov models and selected applications in speech recognition,''
%   \textit{Proceedings of the IEEE}, vol.~77, no.~2, pp.~257-286, 1989.
% \bibitem[3]{Hastie09-TEO}
%   T.\ Hastie, R.\ Tibshirani, and J.\ Friedman,
%   \textit{The Elements of Statistical Learning -- Data Mining, Inference, and Prediction}.
%   New York: Springer, 2009.
% \bibitem[4]{YourName17-XXX}
%   F.\ Lastname1, F.\ Lastname2, and F.\ Lastname3,
%   ``Title of your INTERSPEECH 2021 publication,''
%   in \textit{Interspeech 2021 -- 20\textsuperscript{th} Annual Conference of the International Speech Communication Association, September 15-19, Graz, Austria, Proceedings, Proceedings}, 2020, pp.~100--104.
% \end{thebibliography}
% $$[\mathbf{\Phi_{S_{1}}},\mathbf{\Phi_{N_{1}}},\ldots,\mathbf{\Phi_{S_{3}}},\mathbf{\Phi_{N_{3}}}]$$

\end{document}